\documentclass[twocolumn,floatfix]{revtex4}
\usepackage{graphicx}
\usepackage{amsmath}
\usepackage{amssymb}
\usepackage{bm}

\usepackage{color}

\newcommand{\ket}[1]{|#1\rangle}

\begin{document}

\title{Minimal circuit for a flux-controlled Majorana qubit in a quantum spin-Hall insulator}
\author{B. van Heck}
\affiliation{Instituut-Lorentz, Universiteit Leiden, P.O. Box 9506, 2300 RA Leiden, The Netherlands}
\author{T. Hyart}
\affiliation{Instituut-Lorentz, Universiteit Leiden, P.O. Box 9506, 2300 RA Leiden, The Netherlands}
\author{C. W. J. Beenakker}
\affiliation{Instituut-Lorentz, Universiteit Leiden, P.O. Box 9506, 2300 RA Leiden, The Netherlands}
\date{July 2014}
\begin{abstract}
We construct a minimal circuit, based on the {\em top-transmon\/} design, to rotate a qubit formed out of four Majorana zero-modes at the edge of a two-dimensional topological insulator. Unlike braiding operations, generic rotations have no topological protection, but they do allow for a full characterization of the coherence times of the Majorana qubit. The rotation is controlled by variation of the flux through a pair of split Josephson junctions in a Cooper pair box, without any need to adjust gate voltages. The Rabi oscillations of the Majorana qubit can be monitored via oscillations in the resonance frequency of the microwave cavity that encloses the Cooper pair box.\\
{\tt Contribution for the proceedings of the Nobel Symposium on topological insulators.}
\end{abstract}
\maketitle

\section{Introduction}
\label{intro}

Among the many exotic properties of topological insulators \cite{Has10,Qi11}, the prediction \cite{Fu08} that they can host Majorana zero-modes stands out both for its fundamental interest and for possible applications in topological quantum computing \cite{Nay08}. To braid Majoranas is the prize-winning experiment, since it would identify them as a fundamentally new type of quasiparticles with non-Abelian statistics \cite{Rea00}. The road towards this goal has several milestones, starting from the detection of the zero-mode itself \cite{Mou12,Ali13}.

One intermediate milestone is the construction of a qubit out of Majorana zero-modes and the measurement of its coherence times. This would be essential information for a subsequent braiding experiment to demonstrate its non-Abelian nature. Here we describe a minimal circuit that can initialize, rotate, and read-out the Majorana qubit by coupling it to a transmon (a superconducting charge qubit in a  microwave transmission line resonator \cite{Koc07}). This is the hybrid topological-transmon qubit ({\em top-transmon}) introduced in Ref.\ \onlinecite{Has11}.

The circuit we propose here for the characterization of the Majorana qubit is a reduced version of the full braiding circuit of Ref.\ \onlinecite{Hya13}. By sacrificing the possibility to perform topologically protected operations, we now need only 4 and not 6 Majoranas. For an early generation of experiments this might well be a significant simplification. The reduced circuit shares with the full circuit the feature that all operations are performed by control over Coulomb interactions rather than tunneling \cite{Hec12}. This control is achieved by external variation of magnetic fluxes through macroscopic Josephson junctions, without requiring microscopic control over tunnel couplings.

We focus on Majorana zero-modes induced by the superconducting proximity effect at the edge of a quantum spin-Hall insulator \cite{Fu09}, motivated by recent experimental progress in this direction \cite{Kne12,Du13,Har13}. Relative to the nanowire realization \cite{Lut10,Ore10}, this system has several favorable properties (single-mode conduction, insensitivity to disorder). It also brings along some challenges (how to confine the Majoranas, how to make a T-junction), that we propose to overcome along the lines suggested in Ref.\ \onlinecite{Mi13}.

\section{Top-transmon}
\label{toptransmon}

\begin{figure}[tb]
\centerline{\includegraphics[width=0.8\linewidth]{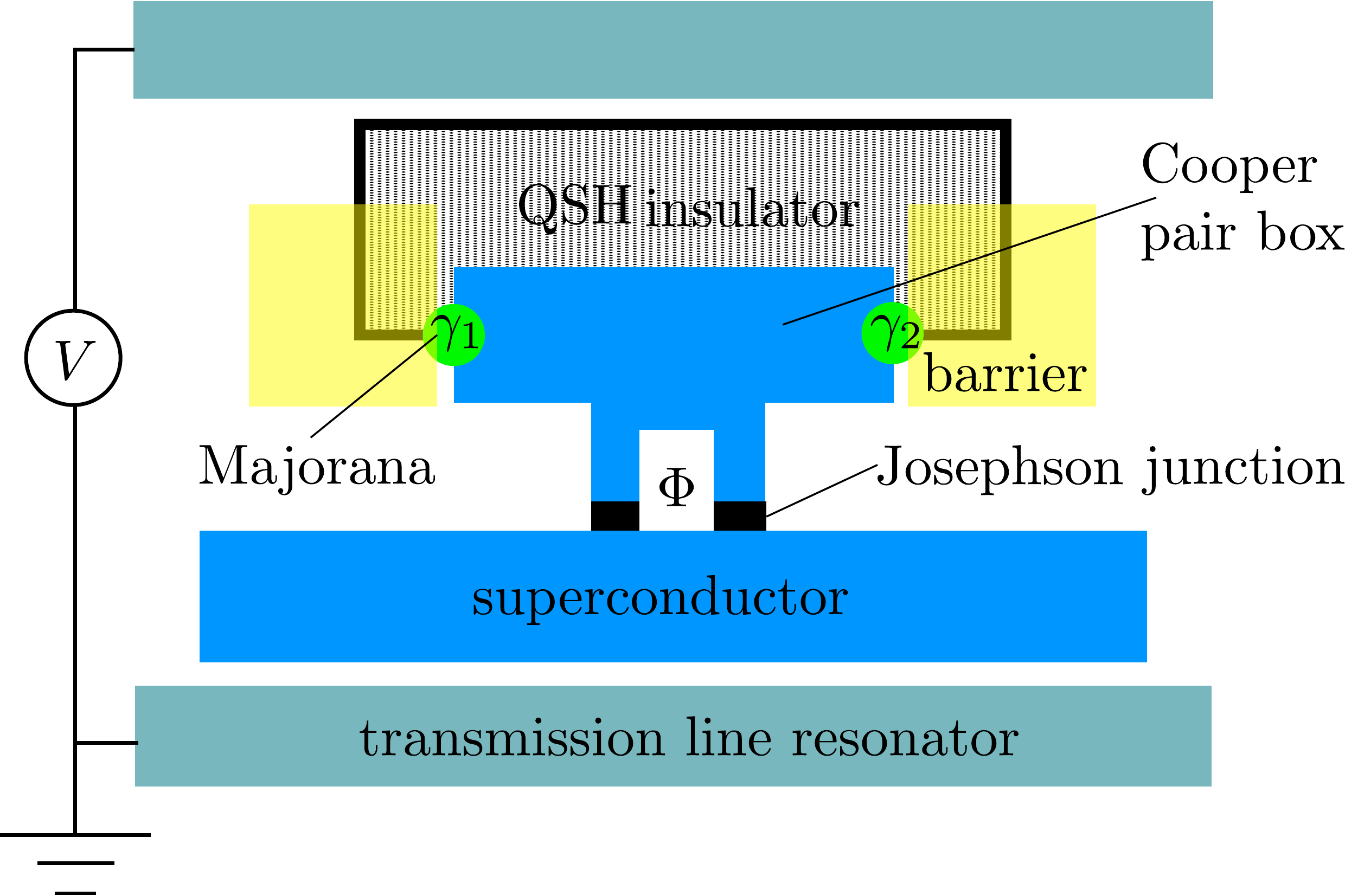}}
\caption{Schematic of a Cooper pair box in a transmission line resonator ({\em transmon}) containing a pair of Majorana zero-modes at the edge of a quantum spin-Hall insulator. This hybrid device ({\em top-transmon}) can couple charge qubit and topological qubit by variation of the flux $\Phi$ through a Josephson junction.
}
\label{fig_transmon}
\end{figure}

Before proceeding to a description in the next section of the minimal circuit that can operate on a Majorana qubit, we summarize the basic ingredients. The device is a hybrid structure \cite{Has11}, dubbed a {\em top-transmon}, combining a topological qubit formed out of Majorana zero-modes with a nontopological transmon qubit.

The basic building block of the transmon, shown in Fig.\ \ref{fig_transmon}, is a Cooper pair box \cite{Mak01} (a superconducting island with charging energy $E_{\rm C}$ $\ll$ Josephson energy $E_{\rm J}$) coupled to a microwave transmission line (coupling energy $\hbar g$). The plasma frequency $\hbar\Omega_{0}\simeq\sqrt{8E_{\rm J}E_{\rm C}}$ is modulated by an amount $\Delta_{+}\cos(\pi q_{\rm ind}/e)$ upon variation of the charge $q_{\rm ind}$ induced on the island by a gate voltage $V$. Additionally, there is a $q_{\rm ind}$-dependent  contribution $\Delta_-\cos(\pi q_{\rm ind}/e)$ to the ground state energy. The charge sensitivity $\Delta_{\pm}\propto\exp(-\sqrt{8E_{\rm J}/E_{\rm C}})$ can be adjusted by varying the flux $\Phi$ enclosed by the Josephson junction, which modulates the Josephson energy $E_{\rm J}\propto\cos(2\pi e\Phi/h)$. In a typical device \cite{Hou09}, a variation of $\Phi$ between $\Phi_{\rm min}\approx 0$ and $\Phi_{\rm max}\lesssim h/4e$ changes $\Delta_\pm$ by several orders of magnitude, so the charge sensitivity can effectively be switched on and off by increasing the flux by half a flux quantum.

Including also the coupling to the microwave photons (creation operator $a^\dagger$ at resonant frequency $\omega_0$), the Hamiltonian of the transmon has the form \cite{Koc07,Hou09}
\begin{align}
H_{\rm transmon}={}&\tfrac{1}{2}\hbar\Omega_0\sigma_{z}+(\Delta_{+}\sigma_z+\Delta_-)\cos(\pi q_{\rm ind}/e)\nonumber\\
&+\hbar\omega_0 a^{\dagger}a+\hbar g(\sigma_{+}a+\sigma_{-}a^{\dagger}).\label{Htransmon}
\end{align}
The charge qubit is represented by Pauli matrices $\sigma_{x},\sigma_{y},\sigma_{z}$, with $\sigma_{\pm}=(\sigma_x\pm i\sigma_{y})/2$.

Majorana zero-modes are represented by identical creation and annihilation operators $\gamma_n=\gamma_n^\dagger$, with anticommutation relation
\begin{equation}
\gamma_n\gamma_m+\gamma_m\gamma_n=2\delta_{nm}.\label{gammadef}
\end{equation}
The number of Majoranas on a superconducting island is necessarily even, say $2N$. They encode a topological quantum number, which is the $\pm 1$ eigenvalue of the fermion parity operator \cite{Kit01}
\begin{equation}
{\cal P}=i^N\prod_{n=1}^{2N}\gamma_n.\label{Pdef}
\end{equation}
The top-transmon Hamiltonian
\begin{align}
H_{\text{top-transmon}}={}&\tfrac{1}{2}\hbar\Omega_0\sigma_{z}+(\Delta_{+}\sigma_z+\Delta_-){\cal P}\cos(\pi q_{\rm ind}/e)\nonumber\\
&+\hbar\omega_0 a^{\dagger}a+\hbar g(\sigma_{+}a+\sigma_{-}a^{\dagger})\label{topHtransmon}
\end{align}
contains a term $\sigma_{z}{\cal P}$ that couples the charge qubit to the topological qubit, see Ref.\ \onlinecite{Hec12} for a derivation.

Since Majorana fermions are charge-neutral particles (being their own antiparticle), one may ask how there can be any coupling at all. The answer is that the state of the $2N$ zero-modes in a superconducting island depends on the parity of the number of electrons on that island, and it is this dependence on the electrical charge modulo $2e$ that provides for a flux-controlled Coulomb coupling between the Majoranas.

A measurement of the resonance frequency $\omega_{\rm eff}$ of the transmission line now becomes a joint projective measurement of the charge qubit and topological qubit \cite{Has11,Hya13},
\begin{equation}
\omega_{\rm eff}=\omega_0+\frac{\sigma_z g^2}{\Omega_0-\omega_0+2{\cal P}\Delta_+/\hbar}.\label{omegaeff}
\end{equation}
This measurement is performed far off resonance ($g\ll |\Omega_0-\omega_0|$, the socalled dispersive regime), so the charge qubit is not excited. If it is in the ground state we may just replace $\sigma_z\mapsto -1$ and $\omega_{\rm eff}$ directly measures ${\cal P}$. In particular, a shift in the resonance frequency signals a bit-flip of the topological qubit.

\section{Minimal circuit}
\label{minimal}

\begin{figure}[tb]
\centerline{\includegraphics[width=0.5\linewidth]{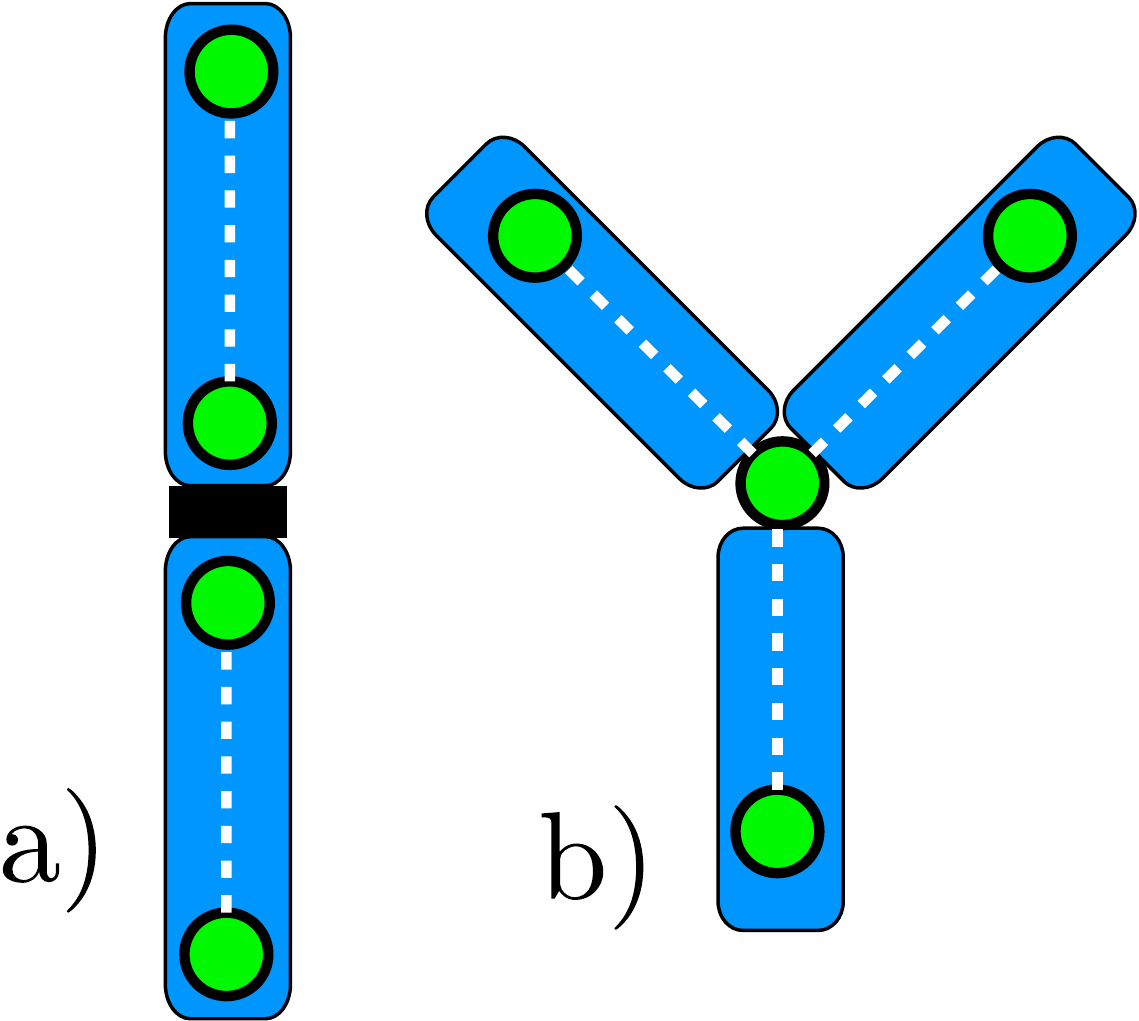}}
\caption{Topological qubit formed out of four Majorana zero-modes, on either two or three superconducting islands. Dashed lines indicate flux-controlled Coulomb couplings, as in the Cooper pair box of Fig.\ \ref{fig_transmon}. In the linear layout (panel a) the coupling between Majoranas on different islands is via a tunnel barrier (thick horizontal line), requiring gate voltage control. By using a tri-junction (panel b) all three couplings can be flux-controlled Coulomb couplings. 
}
\label{fig_layout}
\end{figure}

The conservation of fermion parity on a single superconducting island implies a minimum of two islands for a Majorana qubit, each containing a pair of Majorana zero-modes. The minimal circuit that can operate on a Majorana qubit would then have the linear layout of Fig.\ \ref{fig_layout}a. While the couplings between Majoranas on the same island are flux-controlled Coulomb couplings, the inter-island coupling is via a tunnel barrier, which would require microscopic control by a gate voltage.

\begin{figure}[tb]
\centerline{\includegraphics[width=1\linewidth]{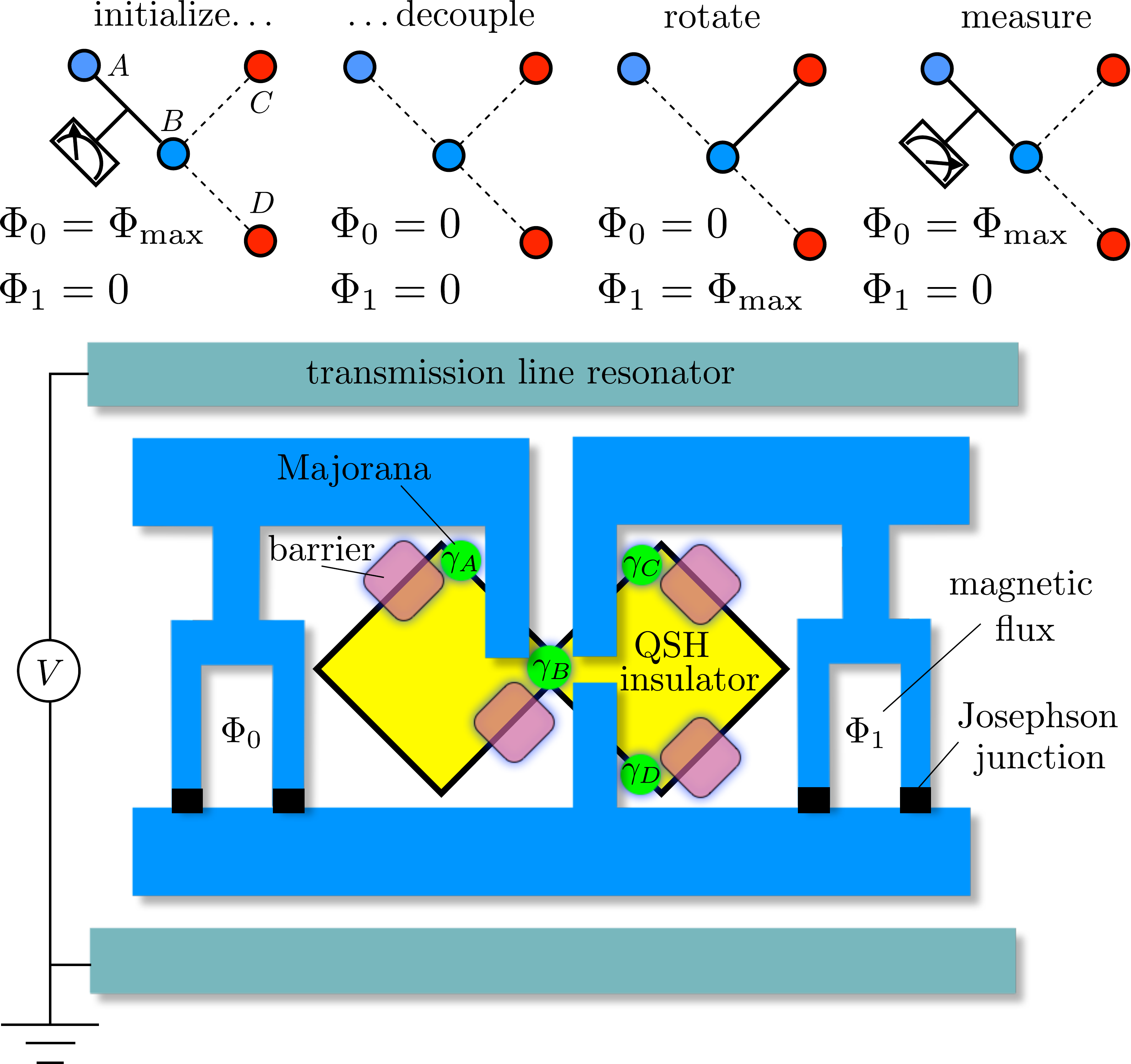}}
\caption{{\em Top-transmon\/} circuit to rotate the qubit formed out of four Majorana zero-modes at the edge of a quantum spin-Hall insulator. One of the Majoranas ($\gamma_B$) is shared by three superconductors at a constriction. The topological qubit is rotated by coupling it to a Cooper pair box in a transmission line resonator ({\em transmon}). The coupling strength is controlled by the magnetic flux $\Phi$ through a pair of split Josephson junctions. The diagrams at the top indicate how the Coulomb couplings of pairs of Majoranas are switched on and off: they are {\em off\/} (solid line) when $\Phi=0$ and {\em on\/}  (dashed line) when $\Phi=\Phi_{\rm max}\lesssim h/4e$. This single-qubit rotation does not have topological protection, it serves to characterize the coherence times of the Majorana qubit.
}
\label{fig_qubit}
\end{figure}

An alternative layout that has only Coulomb couplings needs three rather than two islands, forming a tri-junction as in Fig.\ \ref{fig_layout}b. A tri-junction pins a Majorana zero-mode \cite{Ali11}, which can be Coulomb-coupled to each of the other three Majoranas \cite{Hya13}. The tri-junction also binds higher-lying fermionic modes, separated from the zero mode by an excitation energy $E_{\rm M}$. This is the minimal design for a fully flux-controlled Majorana qubit. In Fig.\ \ref{fig_qubit} we have worked it out in some more detail for the quantum spin-Hall insulator.

Three superconducting islands allow for two independent charge differences, so they produce two charge qubits $\sigma_{z}^{(1)}$ and $\sigma_{z}^{(2)}$. These are coupled to four Majorana zero-modes $\gamma_{A}$, $\gamma_{B}$, $\gamma_{C}$, $\gamma_{D}$. The Hamiltonian is two copies of the top-transmon Hamiltonian \eqref{topHtransmon},
\begin{align}
H={}&\hbar\omega_0 a^{\dagger}a+\sum_{n=1}^{2}\left[\tfrac{1}{2}\hbar\Omega_0^{(n)}\sigma_{z}^{(n)}+\hbar g^{(n)}(\sigma_{+}^{(n)}a+\sigma_{-}^{(n)}a^{\dagger})\right]\nonumber\\
&+i\gamma_A\gamma_B[\sigma_z^{(1)}\Delta_{+}^{(1)}(\Phi_0)+\Delta_{-}^{(1)}(\Phi_0)]\nonumber\\
&+i\gamma_B\gamma_C[\sigma_z^{(2)}\Delta_{+}^{(2)}(\Phi_1)+\Delta_{-}^{(2)}(\Phi_1)],\label{Hcircuit}
\end{align}
where for simplicity we have set $q_{\rm ind}=0$ on each island. We have ignored the higher-lying fermionic modes at the tri-junction, see the Appendix for a calculation that includes these.

Without loss of generality, we will fix the overall parity to be even. The Majorana qubit then has the two states, $|00\rangle$ and $|11\rangle$, in terms of the occupation number of the fermionic modes $c^\dagger_1=\tfrac{1}{2}(\gamma_A+i\gamma_B)$ and $c^\dagger_2=\tfrac{1}{2}(\gamma_C+i\gamma_D)$. Pauli matrices that act on the states ${1\choose 0}=|00\rangle$ and ${0\choose 1}=|11\rangle$ are defined by
\begin{equation}
\tau_x = i\gamma_B\gamma_C,\;\;
\tau_y = i\gamma_A\gamma_C,\;\;
\tau_z = i\gamma_A\gamma_B.\label{taudef}
\end{equation}
With the resonator mode and the charge qubit in their ground state, the Majorana qubit has Hamiltonian
\begin{equation}
H_{\rm M}=\Delta_z(\Phi_0)\,\tau_z +\Delta_x(\Phi_1)\tau_x,\label{Hamqubit}
\end{equation}
with $\Delta_z=\Delta_-^{(1)}-\Delta_+^{(1)}$ and $\Delta_x=\Delta_-^{(2)}-\Delta_+^{(2)}$. Each of the two couplings $\Delta_x(\Phi)$ and $\Delta_z(\Phi)$ can be varied between $\Delta_{\min}$ and $\Delta_{\rm max}$, by variation of the flux between $\Phi_{\rm min}\approx 0$ and $\Phi_{\rm max}\lesssim h/4e$. This circuit does not allow to implement braiding (not enough adjustable couplings). However, it does allow for a complete characterisation of the Majorana qubit.

For starters, one can demonstrate that the four Majoranas constitute a quantum mechanical two-level system, by following these two steps. The first step is the initialization of the qubit in an eigenstate of $\tau_z$, by setting $\Delta_z=\Delta_\textrm{max}$, $\Delta_x=\Delta_{\rm min}$ and waiting for the system to relax to its ground state; or alternatively, one can perform a projective measurement onto a $\tau_z$ eigenstate via microwave irradiation of the transmon qubit \cite{Has11}. Once the qubit is initialised, the second step is to set  $\Delta_x=\Delta_\textrm{max}$. The qubit will then start to rotate around the $x$-axis of the Bloch sphere at a frequency $\Delta_\textrm{max}/\hbar$. This Rabi oscillation can be detected via a shift in the resonant frequency of the microwave transmission line. 

Since the Hamiltonian \eqref{Hamqubit} is that of a fully controllable qubit, and since we are allowed to measure $\tau_z$, all usual qubit tests can be performed. In particular, the coherence times $T_1$ and $T_2$ can be measured. The switching time $T_1$ will likely be dominated by quasiparticle poisoning when all Coulomb couplings are {\em off\/} ($\Delta_x=\Delta_z=\Delta_{\rm min}$). The intrinsic coherence time $T_2$ is usually measured via a Ramsey fringe experiment, applying two $\pi/2$ rotations around the $x$-axis separated by a time delay $\delta t$, while keeping $\Delta_z$ {\em on\/} so that the two qubit states are separated in energy. In the time interval between the $\pi/2$ pulses, the qubit rotates freely around the $z$-axis. A measurement of $\tau_z$ after the second pulse should result in decaying oscillations as a function of $\delta t$, allowing to determine $T_2$. In principle, such measurements can also be used to determine $\Delta_\textrm{min}$ and $\Delta_\textrm{max}$ through the period of the Ramsey fringes.

\begin{figure*}[htb]
\centerline{\includegraphics[width=0.8\linewidth]{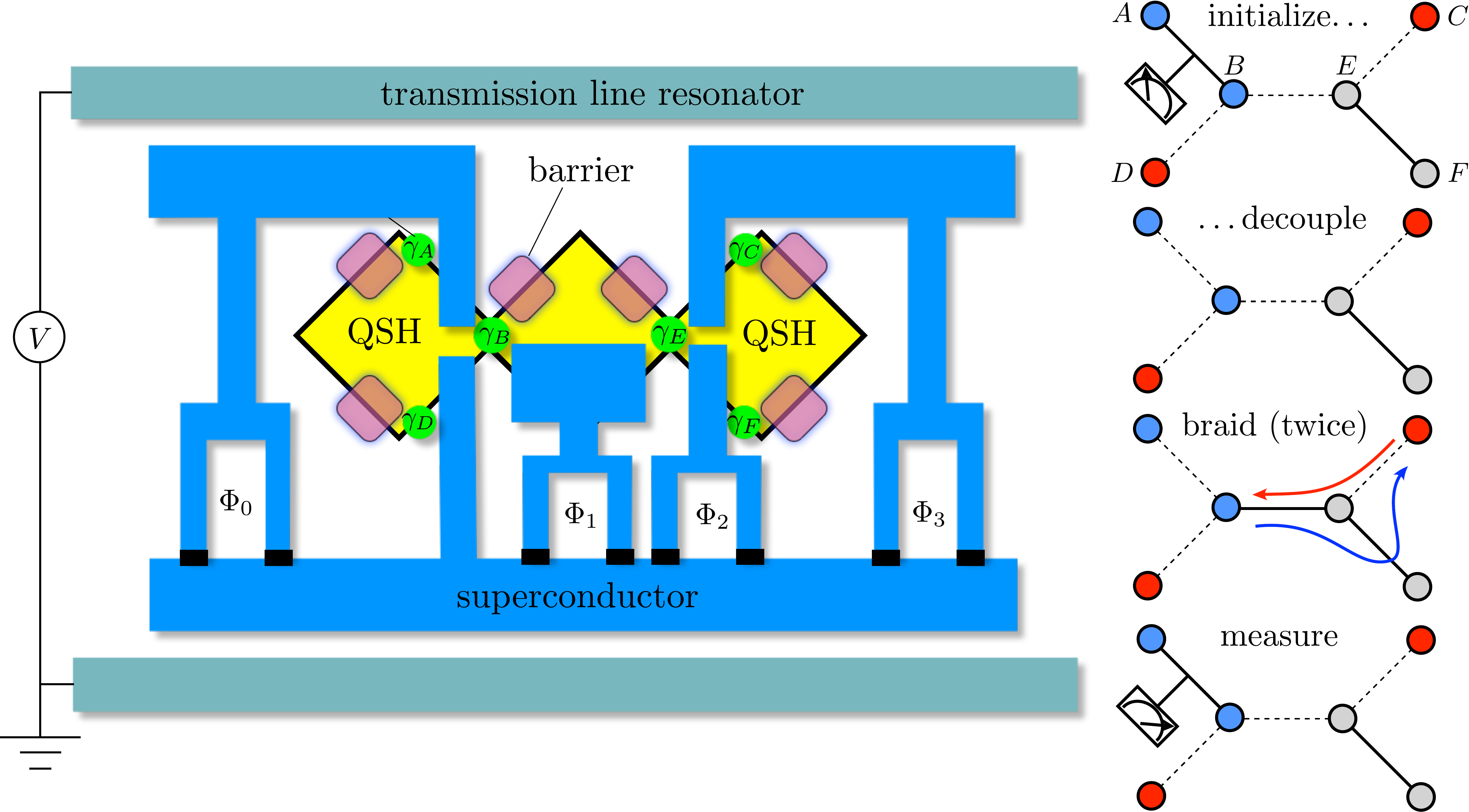}}
\caption{Implementation of the braiding circuit of Ref.\ \onlinecite{Hya13} in a quantum spin-Hall insulator. The two T-junctions are formed by a pair of constrictions. The flux-controlled braiding protocol requires four independently adjustable magnetic fluxes. The Majorana qubit formed out of zero-modes $\gamma_A,\gamma_B,\gamma_C,\gamma_D$ is flipped at the end of the operation, as can be measured via a shift of the resonant microwave frequency. This braiding operation has topological protection.
}
\label{fig_braiding}
\end{figure*}

\section{Characteristic energy scales}
\label{app_energies}

The characteristic energy scales of the two charge qubits are the magnetic flux dependent Josephson energy $E_{\rm J}(\Phi)$ and the charging energy $E_{\rm C}$, which give a plasma frequency $\hbar \Omega_0 \simeq (8 E_{\rm J}E_{\rm C})^{1/2}$. The Josephson and charging energies may or may not be the same on the two islands, that does not matter for the operation of the circuit. 

For the sake of generality we allow for an asymmetry $d$ in the arms of the split Josephson junction, leading to a flux-dependence \cite{Koc07}
\begin{equation}\label{eq_Ej_flux_dependence}
E_{\rm J}(\Phi)=E_{\rm J}^{(0)}\,\cos(e\Phi/\hbar)\,\sqrt{1+d^2\tan^2(e\Phi/\hbar)}.\\
\end{equation}
Typical values of $d$ are in the 10\% range. Hence, for $\Phi_\textrm{max}\simeq h/4e$ one obtains $E_{\rm J}(\Phi_\textrm{max}) \simeq 0.1\, E_{\rm J}^{(0)}$. In the transmon regime one has
\begin{equation}
E_{\rm C}\ll E_{\rm J}(\Phi_{\rm max})\ll E_{\rm J}^{(0)}.\label{ineq2} 
\end{equation}

For a flux-controlled coupling of the Majorana zero-modes we require that the inter-island tunnel coupling $E_{\rm M}$ (across the constriction in Fig.\ \ref{fig_qubit}) and the intra-island Coulomb coupling  satisfy \cite{Hec12}
\begin{equation}
\Delta_{\rm max}, \Delta_{+}(\Phi_{\rm max}) \ll E_{\rm M}\ll E_{\rm J}(\Phi_{\rm max})\ll E_{\rm J}^{(0)}.\label{ineq3}
\end{equation}
The inequalities involving $E_{\rm M}$ should not be interpreted too strictly, in particular since we do not require $E_{\rm M}$ to be under accurate experimental control. In the Appendix we show that $E_{\rm M}$ can vary in a large energy window without compromising the functionality of the device.

 The inequalities can be satisfied for $E_{\rm J}^{(0)} \simeq 300\,{\rm GHz}$, $E_{\rm C}\simeq 5\,{\rm GHz}$, $E_{\rm M}\simeq 5\,{\rm GHz}$ and a split junction asymmetry of $d\simeq 0.1$, such that $E_{\rm J}(\Phi_\textrm{max})\simeq  30$ GHz. Numerical calculation of the energy spectrum for this set of parameters, see Fig.~\ref{fig_spectrum}, yields $\Delta_{\rm max}\simeq120$ MHz, $\Delta_{+}(\Phi_{\rm max})\simeq 0.85$ GHz, and $\Omega_0(\Phi_{\rm max})\simeq 27.5$ GHz, for induced charges close to zero. 

Let us now turn to the parameters of the microwave cavity. The dispersive regime requires $g \ll (\Omega_0\pm 2\Delta_+-\omega_0)$. Furthermore, $g$ should be strong enough that the dispersive frequency shift from Eq.\ \eqref{omegaeff} is large compared to the resonance width $\kappa$,
\begin{equation}
\kappa\ll\omega_{\rm shift}=\frac{4g^2\Delta_{+}(\Phi_{\rm max})}{| \Omega_0(\Phi_{\rm max})-\omega_0|^2-4\Delta_+^2}.\label{measurement}
\end{equation}
Both conditions can be satisfied for  $\omega_0\simeq 25\,{\rm GHz}$, $g\simeq 100\,{\rm MHz}$, $\kappa\simeq 1\,{\rm MHz}$, yielding in particular $\omega_{\rm shift}\simeq 10$ MHz. (We have set $\hbar\equiv 1$.)

The operating temperature should be low enough that excitation of the circuit can be avoided,
\begin{equation}
k_{\rm B}T\ll   E_{\rm M},\; \hbar \Omega_0 ,\;  \Delta_{\rm gap},\label{ineq1}
\end{equation} 
where $\Delta_{\rm gap}$ is the excitation gap induced at the quantum spin Hall edge by the superconducting proximity effect. At $T=10\,{\rm mK}$ the thermal energy $k_{\rm B}T = 1.3\,{\rm GHz}$, so one would need $\Delta_{\rm gap}\gtrsim 10\,{\rm GHz}$.

In the braiding circuit of Ref.\ \onlinecite{Hya13} the initialization of the ancillas also requires that $k_{\rm B}T\ll\Delta_{\rm max}$, so the Coulomb coupling $\Delta_{\rm max}$ cannot be much smaller than $10\,{\rm GHz}$. There is no such requirement for the simpler circuit of Fig.~\ref{fig_qubit}, because no ancillas are needed for the nontopological rotation of a Majorana qubit. This is one reason, in addition to the smaller number of Majoranas, that we propose this circuit for the first generation of experiments on Majorana qubits.

\section{Discussion}
\label{discuss}

The key ingredients of the top-transmon \cite{Has11} are: 1) a charge qubit to couple Majorana zero-modes; 2) a flux-controlled Josephson junction to switch the Coulomb coupling {\em on\/} and {\em off\/}; 3) a microwave resonator to read out the Majorana qubit. There exist many alternative proposals to operate on Majorana qubits \cite{Ali11,Sau11,Fle11,Lei11,Rom12,Hal12,Schmidt13,Pek13,Xue13,Gin13,Che14,Li14,Wan14,Kov14}, including an alternative hybrid design that uses a flux qubit instead of a charge qubit \cite{Hassler10,Sau10,Jia11,Bon11,Mul13,Zha13,Hon13}.

In addition, there is a great variety of candidate systems that could host the Majoranas. Three stand out as being closest to experimental realization: 1) semiconductor nanowires \cite{Mou12,Lut10,Ore10}; 2) chains of magnetic nanoparticles \cite{Cho11,Nad13}; 3) the quantum spin-Hall edge \cite{Fu09,Kne12,Du13,Har13}. All three systems can be integrated with a transmon device, see for example Fig.\ \ref{fig_braiding} for a circuit that can braid the Majoranas via a pair of constrictions in a quantum spin-Hall insulator.

The braiding operation is called ``topologically protected'' because ideally the error is of order $\Delta_{\rm min}/\Delta_{\rm max}$ and can be made exponentially small \cite{Hec12}. Larger errors are to be expected in the first generation of experiments, caused by quasiparticle poisoning \cite{Rainis12}, nonadiabatic effects \cite{Che11, Sch13}, nonequilibrium noise \cite{Chamon11}, and coupling of the Majoranas to localized low-energy states induced by disorder \cite{Cosma}. The quasiparticle poisoning time may well remain as the ultimate limiting factor --- times $\gtrsim 100\,{\rm ms}$ have been reported in Al-Cu devices \cite{Saira}, but the quantum spin-Hall insulator is likely to be less favorable. 

In Figs.\ \ref{fig_qubit} and \ref{fig_braiding} we showed an implementation of the top-transmon circuits at the quantum spin-Hall edge, because of recent experimental developments that suggest this might be a favorable host of Majorana zero-modes \cite{Kne12,Du13,Har13}. The role of T-junctions \cite{Ali11,Hya13}, which in nanowire networks can be fabricated by allowing nanowires to meet and merge during the growth process \cite{Pli13}, is played by constrictions \cite{Mi13}, but since a constriction has four legs rather than three, one of the edges has to be closed off by a barrier. This will require breaking of the time-reversal symmetry that prevents backscattering of the helical edge states \cite{Has10,Qi11}. The weak-field barriers suggested in Ref.\ \onlinecite{Mi13} will presumably not be sufficiently resistive to realize the braiding operation. The alternative is to open up a gap at the edge by a ferromagnetic insulator or by an in-plane magnetic field. Ref.\ \onlinecite{Du13} found no gap opening in their InAs/GaSb quantum wells for in-plane fields up to 10~T, but this might be strongly dependent on the detailed stucture of the quantum wells.

\acknowledgments
This contribution builds on the collective effort of the top-transmon team at Leiden University, as reported in Refs.\ \onlinecite{Has11,Hya13,Hec12}. It  was supported by the Foundation for Fundamental Research on Matter (FOM), the Netherlands Organization for Scientific Research (NWO/OCW), and an ERC Synergy Grant.

\appendix
\section{Energy spectrum of the top-transmon}
\label{appendix}

\begin{figure}[tb]
\centerline{\includegraphics[width=\linewidth]{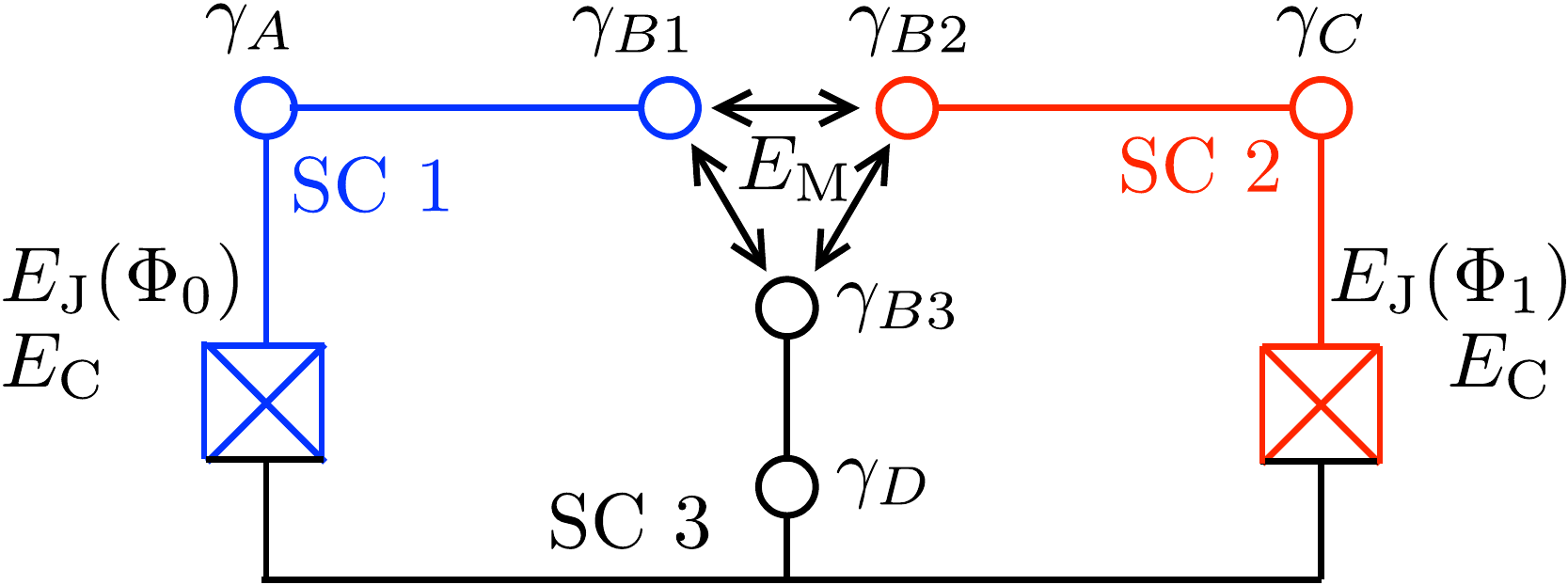}}
\caption{Schematic representation of the top-transmon circuit of Fig.~\ref{fig_qubit}. Colours distinguish different superconducting islands. The three Majoranas coupled by the constriction at the center together produce one zero-mode $\gamma_B$.
}
\label{fig_circuit}
\end{figure}

In the main text we have described the top-transmon circuit of Fig.~\ref{fig_qubit} via the Hamiltonian \eqref{Hcircuit}, which captures the essential features of the coupling of the topological Majorana qubit to the nontopological charge qubit. Two simplifying assumptions are made in this Hamiltonian \cite{Hec12,Hya13}. Firstly, it is assumed that the superconducting phase on each island is pinned to zero by the large Josephson energy $E_{\rm J}\gg E_{\rm C}$, so it does not enter as a dynamical variable. Secondly, the fermionic excited states in the tunnel junction connecting the islands are neglected. In this Appendix we relax both assumptions and calculate the full energy spectrum numerically, following the general procedure of Ref.~\onlinecite{Hya13}. For simplicity we do not include the coupling to the microwave cavity.

\subsection{Full Hamiltonian of the circuit}

A schematic representation of the circuit of Fig.~\ref{fig_qubit} is given in Fig.~\ref{fig_circuit}. The circuit is formed by three superconductors, numbered 1 to 3 in Fig.~\ref{fig_circuit}. Two split Josephson junctions connect the superconductors 1 and 2 to the third one. A further connection between all three superconductors is provided by the quantum spin Hall constriction. We will work in a gauge where all superconducting phases are measured with respect to that of the third superconductor. 

The circuit is described by the Hamiltonian
\begin{equation}
H = H_1 + H_2 + H_{\rm M},\label{HH1H2HM}
\end{equation}
where $H_1$ and $H_2$ are two copies of a Cooper-pair box Hamiltonian describing superconductors 1 and 2,
\begin{equation}
H_n = E_{\rm C}(N_n+q^{(n)}_{\rm ind}/e)^2-E_{\rm J}(\Phi_n)\,\cos(\phi_n-\tilde{\phi}_n).
\end{equation}
The phase and charge operators $\phi_n, N_n$ of the two superconductors are canonically conjugate variables, with commutator $[\phi_n, N_n]=2i$. The charge induced capacitively is $q_\textrm{ind}^{(n)}$. The energy $E_{\rm C}=e^2/2C$ is the charging energy due to the capacitance $C$ to the third superconductor. We have taken the same charging energy for superconductors 1 and 2 and assumed that their mutual capacitance is negligible. The Josephson energies $E_{\rm J}$ of the two Josephson junctions depend on the flux via  Eq.~(\ref{eq_Ej_flux_dependence}). The asymmetry $d_n$ in the arms of each split junction introduces a phase offset $\tilde{\phi}_n$ for each island, determined by $\tan \tilde\phi_n = d_n \tan (e\Phi_n/\hbar)$.

The term $H_{\rm M}$ in Eq.\ \eqref{HH1H2HM} describes the constriction in the quantum spin Hall (QSH) insulator, where three superconducting islands meet. Each superconductor contributes one of the three Majorana modes $\gamma_{B1}$, $\gamma_{B2}$, and $ \gamma_{B3}$. Their tunnel coupling is given by the Hamiltonian
\begin{align}\notag
&H_{\rm M} = iE_{\rm M}  \bigl[\gamma_{B2}\gamma_{B1}\,\cos\,(\tfrac{1}{2}\phi_1-\tfrac{1}{2}\phi_2+\alpha_{12})\nonumber\\
&\;+\gamma_{B1}\gamma_{B3}\,\cos\,(\tfrac{1}{2}\phi_1-\alpha_{13})+\gamma_{B3}\gamma_{B2}\,\cos\,(\tfrac{1}{2}\phi_2+\alpha_{23})\bigr].
\end{align}
We take the same strength $E_{\rm M}$ for all three couplings, but the flux-induced phase shifts differ: $\alpha_{12}=-e(\Phi_0+\Phi_1)/2\hbar$, $\alpha_{23}=e\Phi_1/2\hbar$, and $\alpha_{13}=e\Phi_0/2\hbar$. The three eigenvalues of $H_{\rm M}$ are symmetrically arranged around zero energy, so there is one flux-independent zero-mode. This is the Majorana mode $\gamma_B$ of Fig.~\ref{fig_qubit}. Additionally, there is a fermionic mode at excitation energy $\simeq E_{\rm M}$.

The other Majorana modes of Fig.~\ref{fig_qubit} have no tunnel coupling, so they do not appear explicitly in the Hamiltonian \eqref{HH1H2HM}. They influence the spectrum via a constraint on the number operators  \cite{Fu10},
\begin{equation}
i\gamma_A\gamma_{B1} = (-1)^{N_1},\;\;i\gamma_{B2}\gamma_C = (-1)^{N_2}.\label{constraint1}
\end{equation}
These constraints express the fact that for each island separately the fermion parity (represented on the left-hand-side) equals the number of electrons modulo 2 (represented on the right-hand-side). The product $\gamma_D\gamma_{B3}$ enters only via the global fermion parity of the three superconducting islands, but since this is conserved it does not provide for an independent constraint.

\subsection{Hamiltonian in the measurement configuration}

We wish to extract the parameters $\Omega_0$ and $\Delta_{\pm}$ appearing in Eq.~\eqref{Hcircuit} from the full Hamiltonian \eqref{HH1H2HM}. In order to do so, it is sufficient to consider the measurement configuration of the circuit, i.e.\ set $\Phi_1=0$ and $\Phi_0=\Phi_{\rm max}\simeq h/4e$. The second superconductor then remains in its ground state, and the Hamiltonian reduces to
\begin{align}
H ={}& E_{\rm C}(N_1+q^{(1)}_{\rm ind}/e)^2-E_{\rm J}(\Phi_{\rm max})\,\cos(\phi_1-\tilde{\phi}_1)\nonumber\\
&+iE_{\rm M} \left[\gamma_{B1}(\gamma_{B3}\!-\!\gamma_{B2})\cos\,(\tfrac{1}{2}\phi_1\!-\!\tfrac{1}{4}\pi)+\gamma_{B3}\gamma_{B2}\right].\label{Hfourpi}
\end{align}

For concreteness, we take even global fermion parity,
\begin{equation}
(i\gamma_A\gamma_{B1})\,(i\gamma_{B2}i\gamma_{B3})\,(i\gamma_C\,\gamma_D)=+1.
\end{equation}
The product $i\gamma_C\gamma_D=\pm 1 \equiv P$ is conserved in the measurement configuration, so it can be treated as a c-number. The other products of Majorana operators can be represented by Pauli matrices $\rho_i$,
\begin{subequations}
\begin{align}
&i \gamma_A\gamma_{B1}=P\,i\gamma_{B3}\gamma_{B2}=P\rho_z\,,\\
&i\gamma_{B1}\gamma_{B3}=Pi\gamma_A\gamma_{B2}=P \rho_x\,,\\
&i\gamma_{A}\gamma_{B3}=-Pi\gamma_{B1}\gamma_{B2}=\rho_y\,.
\end{align}
\end{subequations}

Following Ref.\ \onlinecite{Hec11}, we remove the parity constraint \eqref{constraint1} by a unitary transformation,
\begin{equation}
\tilde{H}=U^\dagger H U,\;\;U = \exp\,\left[\frac{i\phi}{4}\,(1-P\rho_z)\right]\,.
\end{equation}
The transformed Hamiltonian is
\begin{align}\notag
\tilde{H}= {}&E_{\rm C}\left[N_1+\tfrac{1}{2}(1-P\rho_z)+q^{(1)}_{\rm ind}/e\right]^2\\\notag
&-E_{\rm J}(\Phi_{\rm max})\cos(\phi_1-\tilde\phi_1)+E_{\rm M} \rho_z\\\notag
&+\tfrac{1}{2}\,E_{\rm M}\,P\,(\rho_x+\rho_y) \big[\cos(\phi_1-\tfrac{1}{4}\pi)+ \cos(\tfrac{1}{4}\pi) \big] \\\label{Htilde}
&+\tfrac{1}{2}\,E_{\rm M}\,(\rho_x-\rho_y)\big[\sin(\phi_1-\tfrac{1}{4}\pi)+\sin(\tfrac{1}{4}\pi)\big]\,.
\end{align}
Notice that, while $H$ from Eq.\ \eqref{Hfourpi} is $4\pi$-periodic in $\phi_1$, the transformed $\tilde{H}$ has become $2\pi$-periodic. This is why now we can forget about the parity constraint \eqref{constraint1} and straightforwardly diagonalize the Hamiltonian.

\subsection{Energy spectrum in the measurement configuration}

\begin{figure*}[tb]
\centerline{\includegraphics[width=\linewidth]{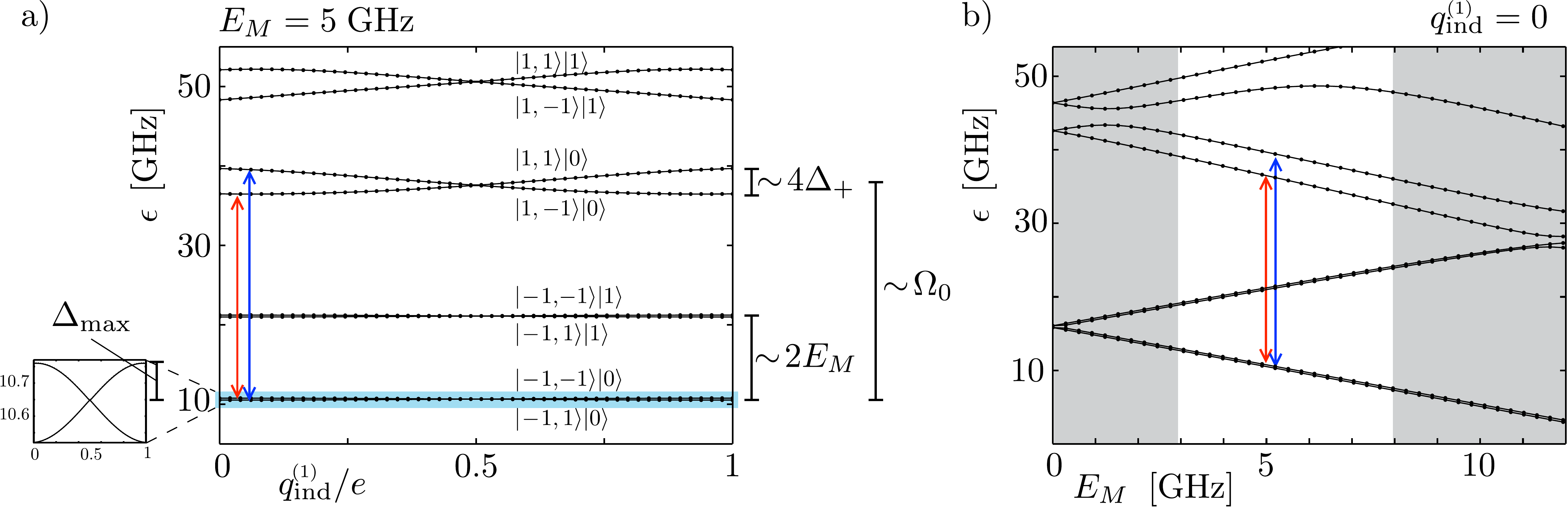}}
\caption{Energy spectrum of the top-transmon circuit of Fig.~\ref{fig_qubit}, obtained from numerical diagonalization of the Hamiltonian \eqref{Htilde} for $E_{\rm J}=300$ GHz, $E_{\rm C}=5$ GHz, $\Phi_\textrm{max}=h/4e$. The junction asymmetry was $d=0.1$, so that $E_{\rm J}(\Phi_{\rm max})\simeq 30$ GHz. In panel (a), the lowest eight energy levels for $E_{\rm M}=5$ GHz are shown as a function of the induced charge $q^{(1)}_{\rm ind}$. They correspond to the eight eigenstates $\ket{\sigma,\tau}\ket{f}$, where $\sigma=\pm1$ labels the excited/ground state of the charge qubit, $\tau=\pm 1$ labels the even/odd parity state of the topological qubit, and $f=0,1$ the occupation number of the fermionic  state in the constriction. As indicated by the colored arrows, the ground and excited state of the charge qubit are separated by an energy $\Omega_0\pm2\Delta_+ \simeq (27.5)\pm(1.7)$ GHz, depending on the state of the topological qubit. The inset shows the weak charge dispersion of the ground state doublet ($\Delta_{\rm max}\simeq 120$ MHz). In panel (b), the same energy levels are shown as a function of the tunnel coupling $E_{\rm M}$ for a fixed value of $q^{(1)}_\textrm{ind}=0$. For a proper operation of the circuit it is required that the states $f=1$ with an excited fermionic mode are well separated from both ground and excited states of the charge qubit. We have highlighted between grey panels a large energy window $3\; {\rm GHz} \lesssim E_{\rm M} \lesssim 8\; {\rm GHz}$ where this requirement is met.}
\label{fig_spectrum}
\end{figure*}

We numerically diagonalize the Hamiltonian $\tilde{H}$ in the basis of eigenstates of $N_1$ and $\rho_z$, truncating the Hilbert space until convergence is reached. To obtain the full spectrum for even global fermion parity, we diagonalise $\tilde{H}$ for both values of $P=\pm 1$ and merge the results. The low-lying part of the spectrum is shown in Fig.~\ref{fig_spectrum} for the choice of parameters of Sec.~\ref{app_energies}.

From the effective Hamiltonian~\eqref{Hcircuit}, we can identify two good quantum numbers for the low-lying part of the spectrum of $\tilde{H}$ in the measurement configuration: the $\sigma_z$ eigenvalues $\sigma=\pm 1$ of the charge qubit and the $\tau_z$ eigenvalues $\tau=\pm 1$ of the topological qubit. Additionally, there is the occupation number $f=0,1$ of the fermionic state in the constriction. These three quantum numbers can be used to label the eight lowest energy states $\ket{\sigma, \tau}\ket{f}$ and their energies $\epsilon^{f}_{\sigma,\tau}$. The top-transmon parameters $\Omega_0$, $\Delta_{\pm}$, and $\Delta_{\rm max}$ follow from
\begin{subequations}
\begin{align}
\Omega_0&=\tfrac{1}{2}[(\epsilon^{0}_{+1,+1}+\epsilon^{0}_{+1,-1})-(\epsilon^{0}_{-1,-1}+\epsilon^{0}_{-1,+1})]\\
\Delta_{\pm}&=\tfrac{1}{4}[(\epsilon^{0}_{+1,+1}-\epsilon^{0}_{+1,-1})\pm(\epsilon^{0}_{-1,-1}-\epsilon^{0}_{-1,+1})]\\
\Delta_{\rm max} &= \Delta_+-\Delta_-=\tfrac{1}{2}(\epsilon^{0}_{-1,-1}-\epsilon^{0}_{-1,+1})\,.
\end{align}
\end{subequations}

\end{document}